# THEORETICAL MODELING AND SIMULATION OF PHASE–LOCKED LOOP (PLL) FOR CLOCK DATA RECOVERY (CDR)

**Z. M. ASHARI AND A.N. NORDIN**

*Electrical and Computer Engineering Department,*
*Kulliyyah of Engineering, International Islamic University Malaysia,*
*Jalan Gombak, 53100 Kuala Lumpur, Malaysia.*

*zainab_ashari@yahoo.com*

***ABSTRACT:*** Modern communication and computer systems require rapid (Gbps), efficient and large bandwidth data transfers. Agressive scaling of digital integrated systems allow buses and communication controller circuits to be integrated with the microprocessor on the same chip. The Peripheral Component Interconnect Express (PCIe) protocol handles all communcation between the central processing unit (CPU) and hardware devices. PCIe buses require efficient clock data recovery circuits (CDR) to recover clock signals embedded in data during transmission. This paper describes the theoretical modeling and simulation of a phase-locked loop (PLL) used in a CDR circuit. A simple PLL architecture for a 5 GHz CDR circuit is proposed and elaborated in this work. Simulations were carried out using a Hardware Description Language, Verilog-AMS. The effect of jitter on the proposed design is also simulated and evaluated in this work. It was found that the proposed design is robust against both input and VCO jitter.

***ABSTRAK:*** Sistem komunikasi dan komputer moden memerlukan pemindahan data yang cekap (Gbps), dan bandwidth yang besar. Pengecilan agresif menggunakan teknik sistem digital bersepadu membenarkan bas dan litar pengawal komunikasi disatukan dengan mikroprocessor dalam cip yang sama. Protokol persisian komponen sambung tara ekspres (PCIe) mengendalikan semua komunikasi antara unit pemprosesan pusat (CPU) dan peranti perkakasan. Bas PCIe memerlukan litar jam pemulihan data (CDR) yang cekap untuk mendapatkan kembali isyarat jam yang tertanam dalam data semasa transmisi. Karya ini menerangkan teori pemodelan dan simulasi gelung fasa terkunci (PLL) untuk CDR. Rekabentuk 5 GHz PLL yang mudah telah dicadangkan dalm kertas kerja ini. Simulasi telah dijalankan menggunakan perisian verilog-AMS. Simulasi mengunnakan kesan ketar dalam reka bentuk yang dicadangkan telah dinilai. Reka bentuk yang dicadangkan terbukti teguh mengatasi ganguan ketar di input dan VCO.

## 1. INTRODUCTION

Aggressive scaling in accordance with Moore’s Law has allowed miniaturization of both the processor and its communication controller circuits. Submicron transistors result in CPUs with GHz clock speeds. High speed data transfer rates (Gbps) between peripheral components and the central processing unit (CPU) is important to avoid bottlenecks in modern computer systems. The Peripheral Component Interconnect Express (PCIe) protocol handles all communications between the hardware devices and the CPU. During transmission, clocks are often embedded in the data to allow efficient synchronization between the transmitter and receiver modules. As such, the Clock Data Recovery (CDR) circuit is an important component in a PCIe system. This scheme is vulnerable to jitter, clock skews and channel latency. The quality of the clock is difficult to optimize with increased bandwidth demands and high speed transfer rates. Different clock architectures can be used for synthesis, distribution and recovery of I/O clocks. To cater for high-speed data links different techniques can be used namely: clock multiplication, forwarded clock recovery, embedded clock recovery, per-pin deskew, jitter filtering and duty-cycle error correction [1].

The first generation of I/O interconnects known as PCIe 1.0, had initial speeds of 2.5 Gbps. The next generation, PCIe 2.0 is based on PCIe 1.0 principles, but it supports speeds of 5Gbps. The ability of the CDR circuit to influence PCIe's performance during Gbps data transmission makes its design critical. There are different CDR circuit top such as delay-locked loop (DLL) and phase-locked loop (PLL). The DLL is a feedback system, similar to PLL but the DLL tracks the phase of a reference signal [2]. The DLL circuit is a simple, robust and stable clock generator [1]. Unlike a PLL, DLL do not filter input reference jitter. PLL is generally known as a negative-feedback system with forward gain term and feedback term. It purposely used to synchronize the output frequency by an oscillator with the reference signal [3] and works as a demodulator [4]. In other words, it generates a signal with less phase difference. PLLs based are often implemented in designing CDR because they offer filtering to the phase or frequency signal. With transfer rate of 5 Gbps, PLL approached are preferably chosen in [5] and [6].

Theoretical modeling and simulation of high performance of CDR based on the PLL transfer functions is demonstrated in this work. Modeling of analog mixed-signal architecture for dedicated CDR circuit using Verilog-AMS HDL is proposed. Six important sections are presented in this paper. The system architecture of PLL is described in Section2. Section 3 illustrates building blocks and transfer function of proposed CDR circuit. Jitter in PLL, simulation and results, discussion, and conclusion are analyzed in section 4, 5 and 6.

## 2. PROPOSED CDR ARCHITECTURE

The PLL comprises of a phase detector (PD), a low-pass filter, voltage controlled oscillator (VCO) and a feedback divider, as illustrated in Fig. 1. This PLL is designed to generate high output frequency of 5 GHz from a low frequency input of 50 MHz. The input source is derived from the embedded clock in the data stream. This PLL then compares this input at the PD with the reference clock generated from the divider circuit. The phase difference between the reference clock and the input source feeds the loop filter. The loop filter produces a DC voltage that controls the VCO. The VCO generates a high frequency output (GHz) which is also the recovered clock. In the proceeding analysis, each block was using its transfer function.

### 2.1 Phase detector (PD)

Phase detector (PD) is the first block in PLL architecture. It is used to compare the phases of the incoming data with the phase of the clock generated by the divider circuit [7]. The phase detector determines the performance of a CDR circuit. The best phase detector basically should accomplish three essential functions which are data transition detection, phase difference detection and have low spur noise [8]. Phase difference between the input phase and the output phase as well as the phase difference between the input voltage and the output voltage are also of concern.

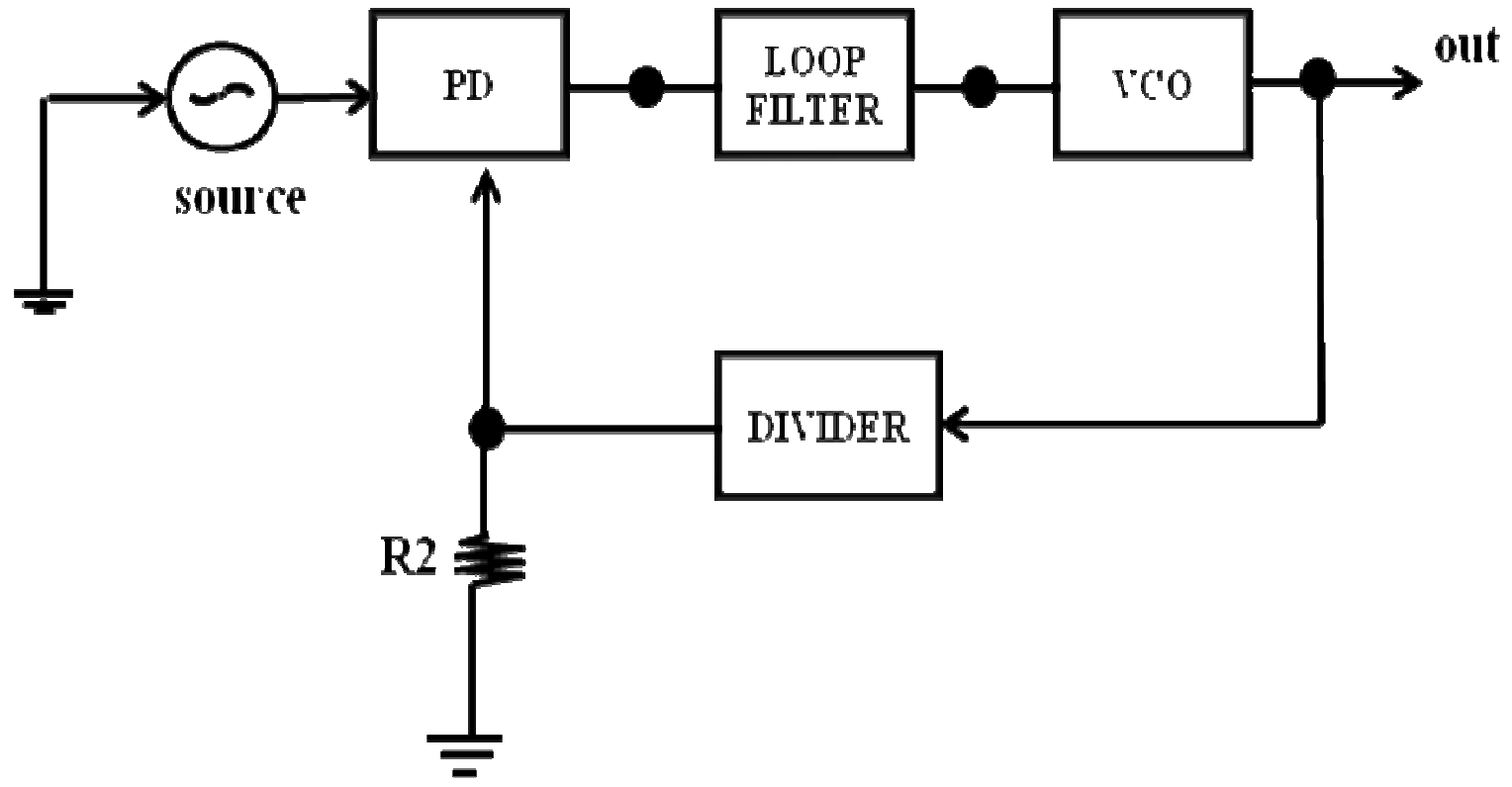

Fig. 1: Circuit design of proposed PLL.

In locked condition, the phase difference between the output phase, $\phi_{out}$ and the input phase, $\phi_{in}$ should be small and constant. It makes the output angular frequency, $\omega_{out}$ equal to angular frequency of the input, $\omega_{in}$. The output voltage of the PD, $V_{PD}$ is proportional to the phase difference of both inputs. The value of $V_{PD}$ can be calculated according to the equation (1) below:

$$V_{PD} = K_{PD}(\phi_{out} - \phi_{in}) = K_{PD}\Delta\phi \quad (1)$$

where $K_{PD}$ is a PD gain and $\Delta\phi$ is a phase error between two inputs. The $K_{PD}$ also can be determined based on the slope of $V_{PD}$ and $\Delta\phi$. The width of the output pulses at PD varies if the phase difference of the two inputs is also varied. In order to reduce the phase error in a PLL system, $K_{PD}$ and $K_{VCO}$ must be large.

## 2.2 RLC Low-Pass Filter

The following block in the PLL system after a phase detector is a loop filter. A low pass filter is applied as a loop filter to pass lower frequency output of the phase detector and suppress the higher frequency jitters. In most CDR circuit designs, a simple RLC low pass filter with one pole and one zero is frequently used [8, 9]. The transient response of a stable low-pass filter depends on the magnitude of the pole and zero. The number of poles and zeroes in low pass filter verify the type of PLL system.

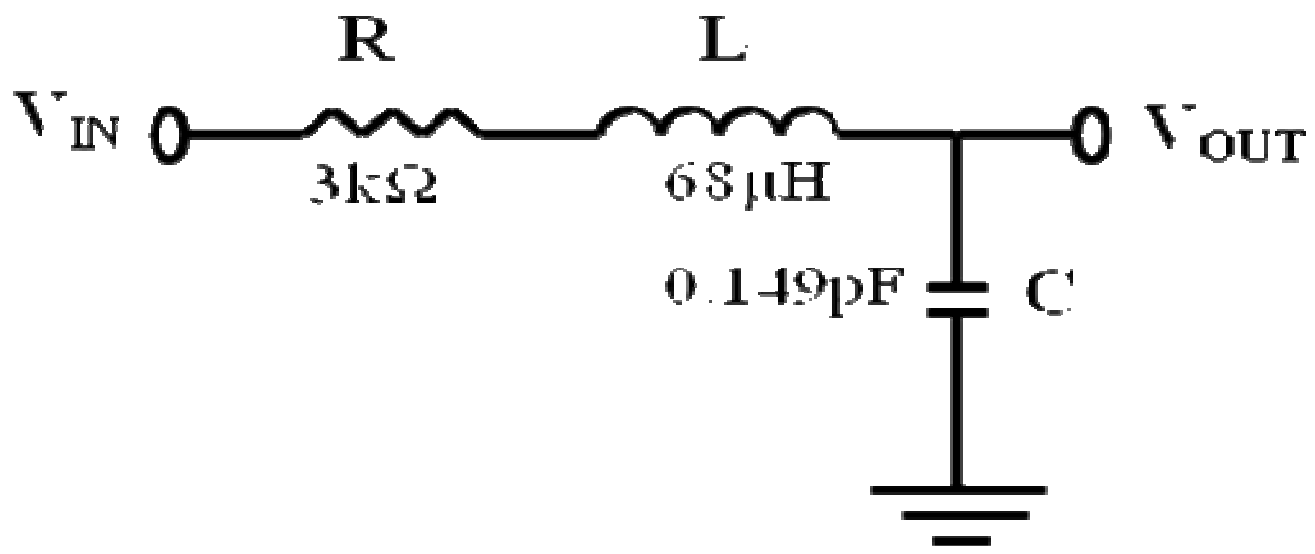

Fig. 2: Low-pass filter circuit.

$$X_L = j2\pi fL \quad (2)$$

$$X_C = \frac{1}{j2\pi fC} \tag{3}$$

$$Z = \sqrt{(X_L - X_C)^2 + R^2} \tag{4}$$

In PLL applications, loop filters of second order and higher are can be applied to offer better noise filtering. Figure 5 shows RLC low pass filter circuit. The circuit design and numerical values of RLC low pass filter is verified according to the following equations (2) to (4). The inductor's impedance, $X_L$, capacitor's impedance, $X_C$, and total impedance of the loop filter circuit, *Z* were determined.

The function of low pass filter is to diminish jitter, which is generated at the input of PD. Jitter exists as the input data stream experiences pulses or ripples. The RLC low pass filter transfer function is identified as shown below from (5) to (9). Equation (5) shows a second order transfer function of RLC low pass filter. The equation (6) is the same as the equation (5) except that the denominator has been organized with $a_1 = 1/LC$ and $a_2 = R/L$ so highlight that the denominator is a quadratic equation. The actual factoring processes for the gain is based on denominator in (6) with values of *s* are given by $s_1$ and $s_2$. The quadratic equation is used to determine the value of $s_1$ and $s_2$ as illustrated in equation (7) and (8). The two poles are complex and the resulting transfer function of $G(s) = 0.4672+5.02X10^{-11}j$ was obtained based on (9) and the calculated R, L, C values are shown in Fig. 2.

$$G(s) = \frac{1/LC}{s^2 + R/L\,s + 1/LC} \tag{5}$$

$$G(s) = \frac{a_1}{s^2 + a_2 s + a_1} \tag{6}$$

$$s_1 = -\frac{a_2 + \sqrt{a_2^2 - 4a_1}}{2} \tag{7}$$

$$s_2 = -\frac{a_2 - \sqrt{a_2^2 - 4a_1}}{2} \tag{8}$$

$$G(s) = \frac{1/LCs_1 s_2}{\left(1 - \frac{s}{s_1}\right)\left(1 - \frac{s}{s_2}\right)} \tag{9}$$

## 2.3 VCO

Another important part of the PLL is the VCO. The major function of VCO is to create an oscillating signal proportional to the applied control voltage [10]. According to [11], the value of $V_{VCO}$ depends on the DC component of $V_{cont}$ which extracted from $V_{PD}$ via low-pass filter. There are two unknown quantities, which need to be calculated for the VCO to function properly, namely phase error, $\phi_{err}$ and $V_{cont}$ (controlled voltage). The $V_{cont}$ will be equal to $V_{out}$ as the input and output frequencies are equal to $\omega_{out}$. Thus, the VCO and PD characteristics required to make it reliable are derived as (10) to (13):

$$\omega_{out} = \omega_{in} + K_{vco} V_{cont} \tag{10}$$

$$\varphi_{err} = \frac{V_{cont}}{K_{PD}} = \frac{\omega_{out} - \omega_{in}}{K_{PD} K_{vco}} \tag{11}$$

$$G_{VCO}(s) = \frac{K_{VCO}}{s} \tag{12}$$

Based on (10), output frequency of the VCO, $\omega_{out}$ is depends on $\omega_{in}$ (input frequency signal), $K_{VCO}$ (VCO gain) and the DC output of the loop filter, $V_{cont}$. $V_{cont}$ in (11) can also be used to describe the difference between input and output frequency signal. When $V_{cont}$ is divided over the VCO gain, $\phi_{err}$ is obtained. The transfer function of the VCO block is demonstrated in (12) which consist of $K_{VCO}$. Two important characteristics are recognized after the loop returns to lock. All parameters such as the voltage due to the phase difference, $V_{PD}$, and VCO frequency remain constant except for total input and output phases. Otherwise, $V_{cont}$ from the output low pass filter can be utilized in analyzing PLL system [11].

## 3. SIMULATIONS OF JITTER IN PLL

The existence of jitter is normally happens during the transmission and receive process. This phenomenon is indirectly effects overall system performance in PLL. The transmission signal or data will experience noise and distortion especially for large bandwidth data. As a result, transfer rate and speed of a PLL system decreased. Small changes in period during transmission and receive process affects the excess phase and the total phase of the waveform. In other words, excess phase and total phase varies as the the waveform becomes unstable.

Jitter can be categorized according to its speed as fast jitter or slow jitter. Slow jitter occurs whenever frequency of a signal varies slowly from one period to the other whereas fast jitter is due to changes in period. Trend of jitter in PLL can be modelled as an input to the PD, typically known as the input excess phase ($\Phi_{in}$) and as a random component ($\Phi_{VCO}$) at the VCO. Generally, jitter is generated in PD and finally produced at the VCO output. In fact, it can also be purposely injected at PD and VCO circuit to test the ability of a system in withstanding distortion signals as demonstrated in Fig. 3 and Fig. 6. The SMASH Dolphin Integration Verilog-AMS tools have been used for all the simulation presented in this section. All the results in are based on the effect of injected jitter in input PD and VCO.

### 3.1 Jitter at Input PD

CDR's circuit performance in sensing and recovering distortion signal at the input is tested in this section. Figure 3 shows that jitter is added at the input of PD. The distortion signal is superimposed on the 50 MHz input and passes through the PD, loop filter, VCO and divider. The resulting output 5 GHz waveform ($\Phi_{out}$) is not affected by the jitter at input $\Phi_{in} > 0$ as verified in Fig. 5.

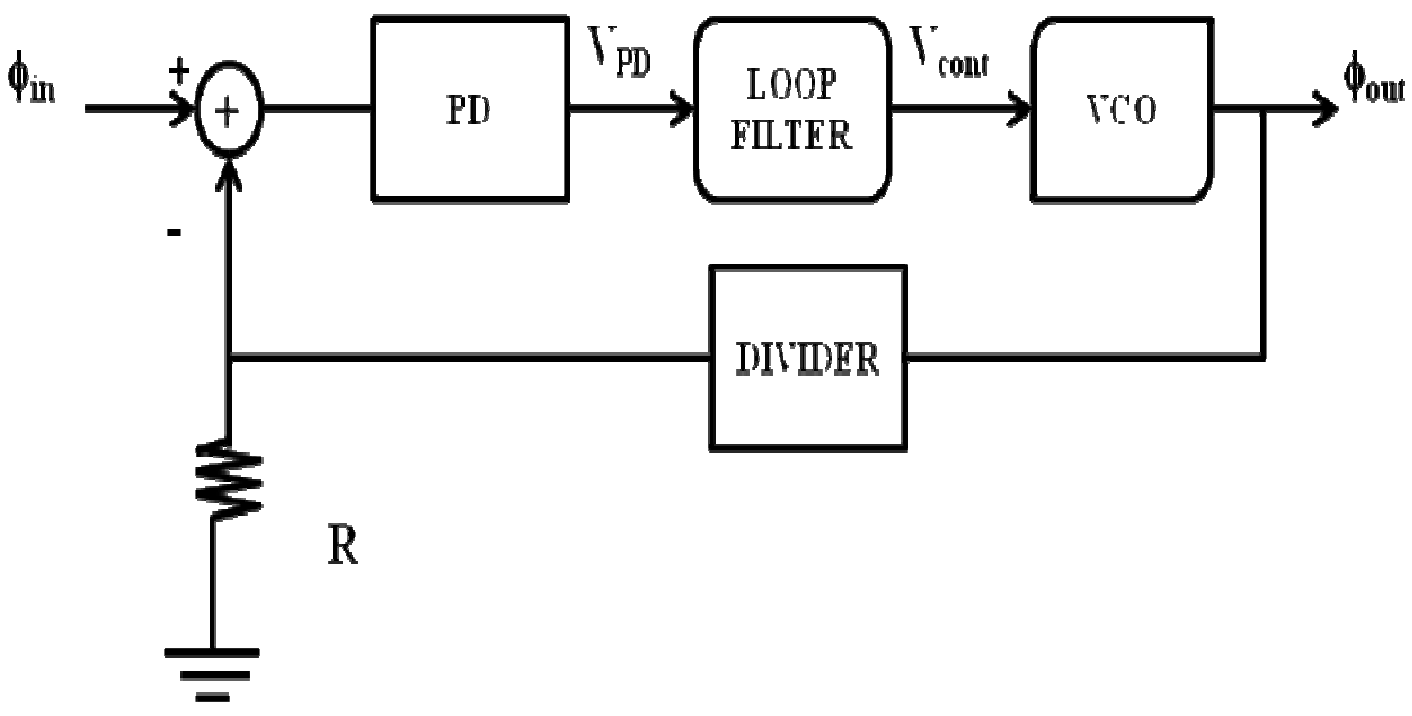

Fig. 3: Effect of PD jitter.

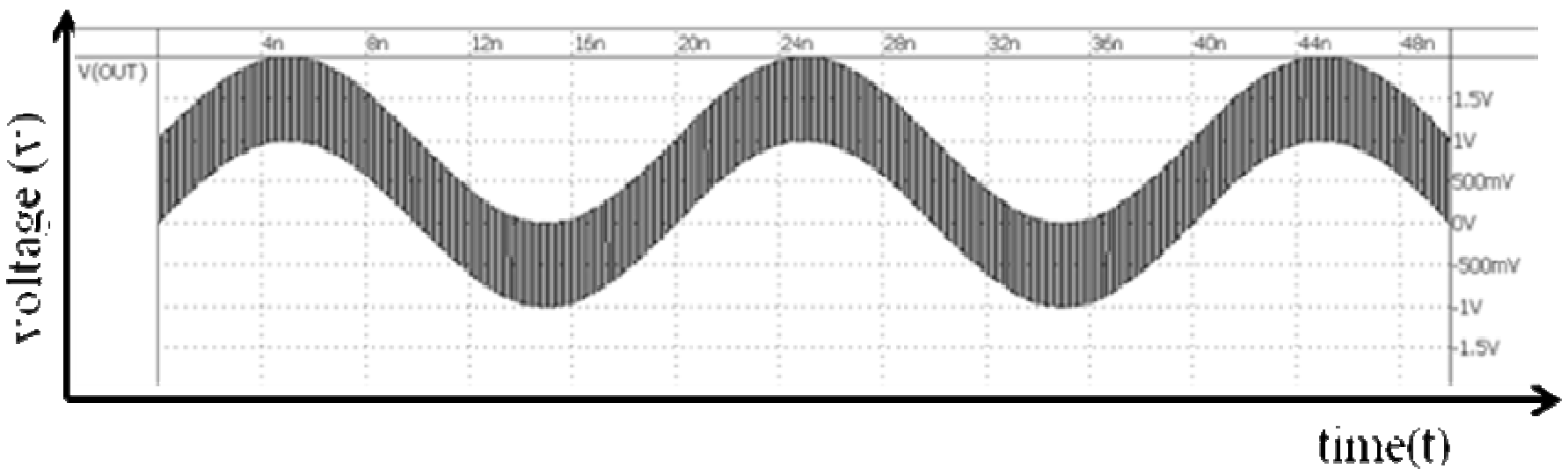

Fig. 4: Input signal of 50 MHz with jitter.

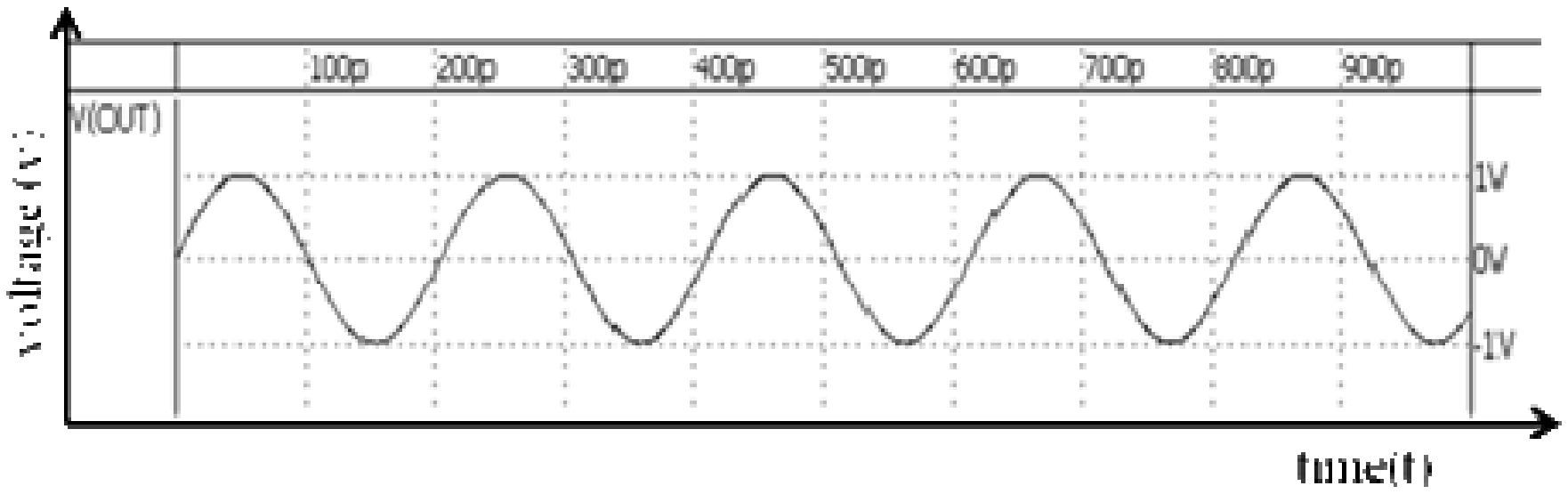

Fig. 5: Output waveform of proposed PLL 5 GHz ($\Phi_{out}$).

## 3.2 Jitter at VCO

In this part, jitter is injected at the VCO block, ($\Phi_{VCO}$) and no jitter is applied at the input PD, $\Phi_{out} = 0$. At PD, two 50 MHz input signals are compared as shown in Fig. 7 and Fig. 8. A clean output signal of 5 GHz is generated after passing through PD, loop filter and VCO. The distortion signal shown in Fig. 9 is an input to the VCO. The VCO shows robust noise characteristics as the output of the PLL produces a clean signal as shown in Fig. 10.

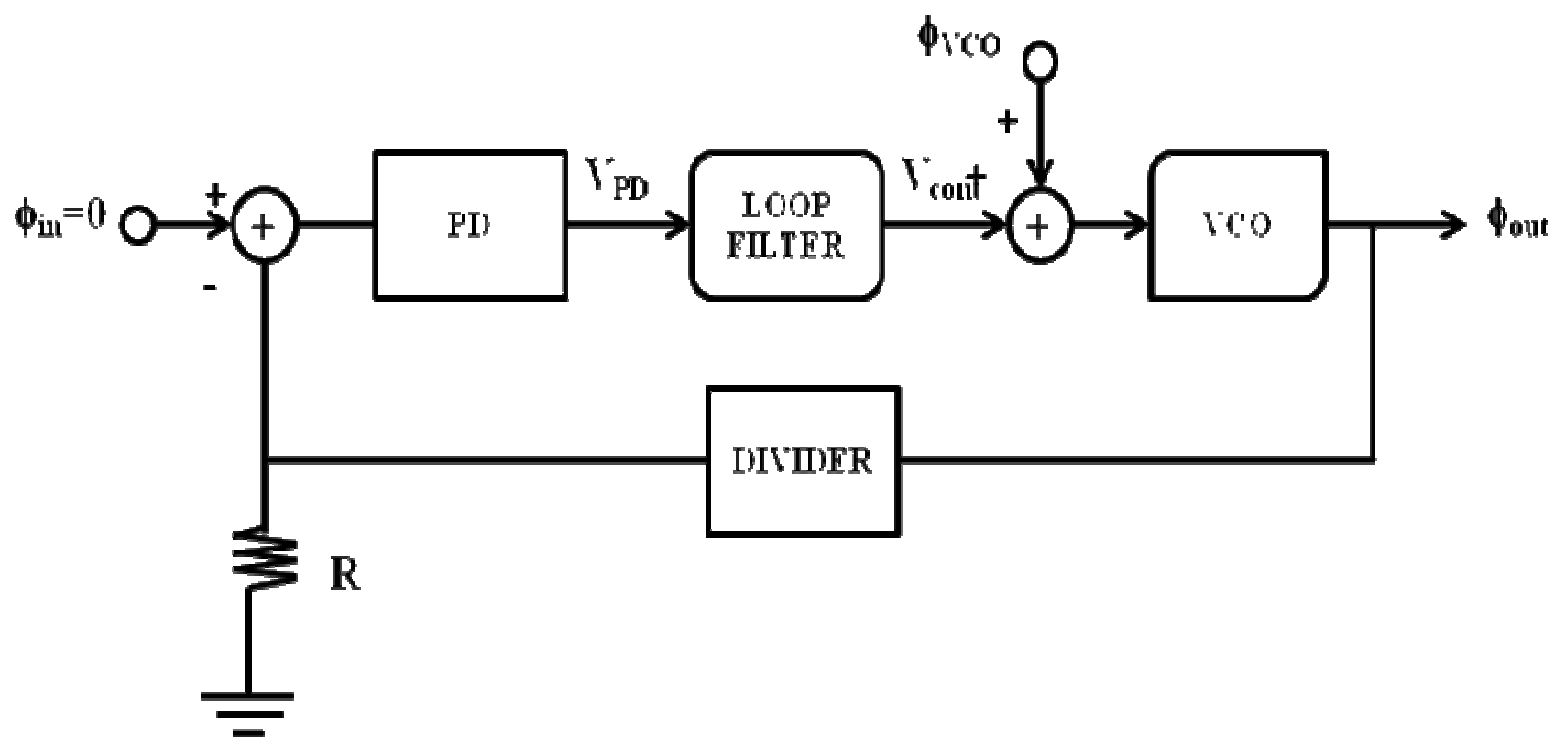

Fig. 6: Effect of VCO jitter.

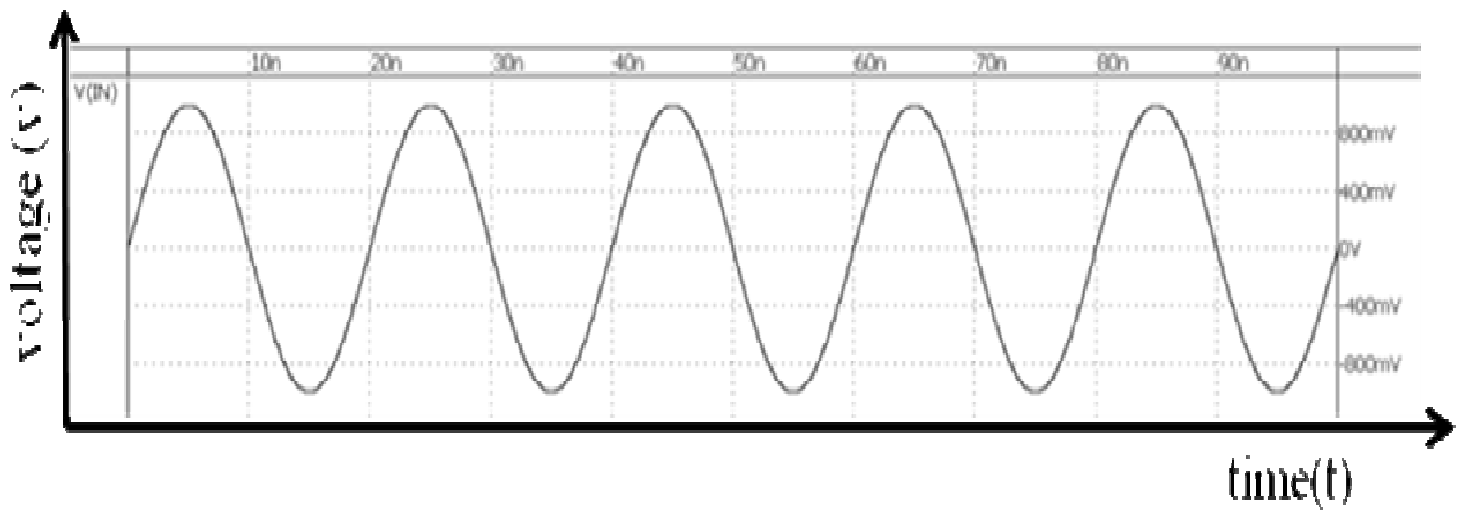

Fig. 7: Input signal of 50 MHz without jitter.

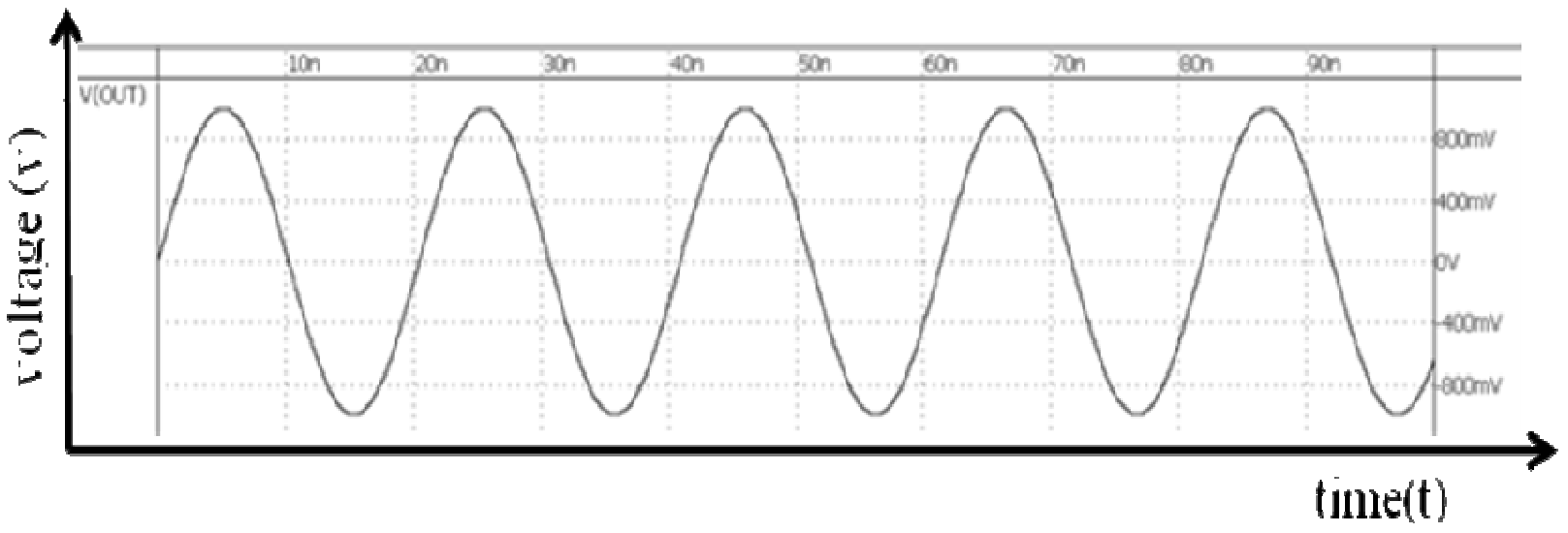

Fig. 8: Feedback waveform of the divider.

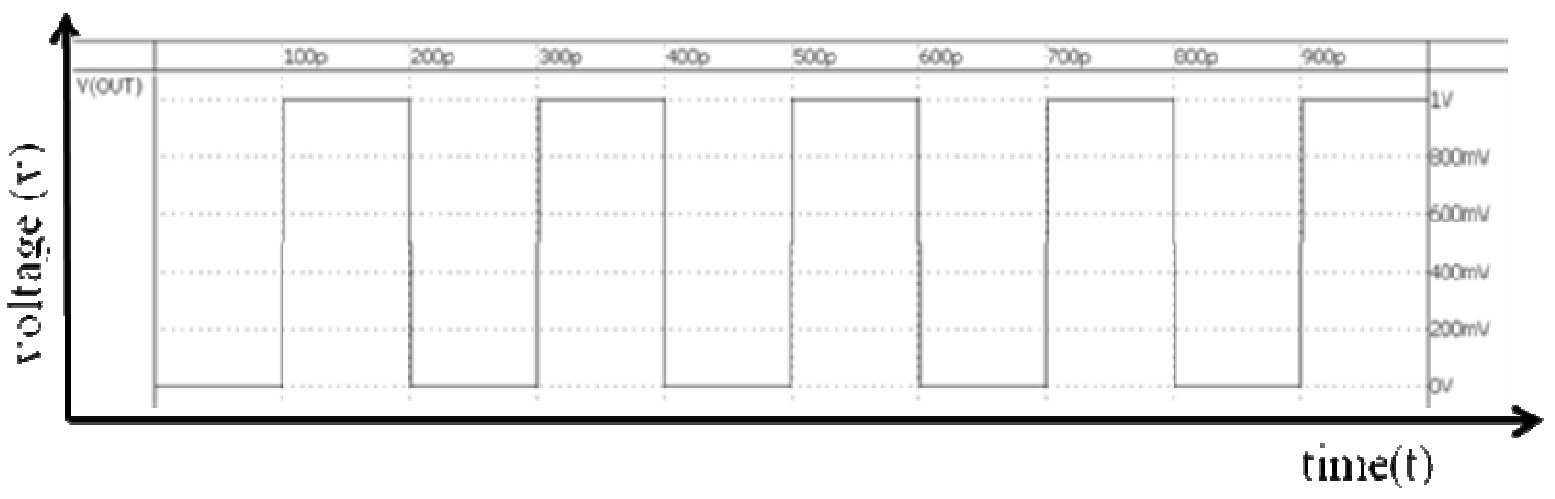

Fig. 9: Waveform signal of jitter, ($\Phi_{VCO}$).

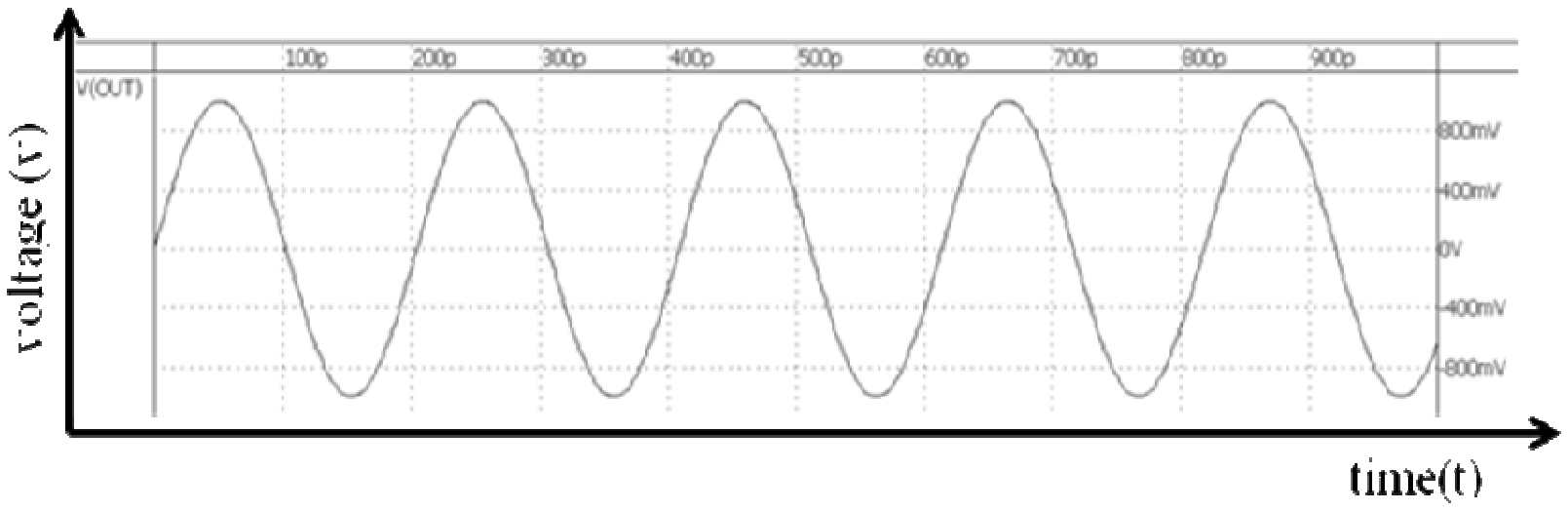

Fig. 10: Output waveform of proposed PLL 5 GHz ($\Phi_{out}$).

## 4. DISCUSSION

This work proposes a design of a 5 GHz CDR PLL circuit comprising of an RLC loop filter, PD, VCO and divider circuit connected in feedback. The PD compares the clock embedded in the data stream with a reference clock. Its output is sent to an RLC circuit. The RLC low-pass filter is used to ensure the CDR's stability by filtering high frequency noise. The RLC is connected to a VCO, which can generate a stable high frequency sinusoidal signal based on the DC voltage output from the RLC. The divider block divides the output frequency produced by the VCO by a factor of N in order to be equal to the input frequency signal of 50 MHz.

The effects of jitter are illustrated using simulation results. Jitter, ($\Phi_{in}$ ) is purposely injected at the input PD to test CDR circuit performance as shown in Fig. 3. Based on the resulting waveform in Fig. 5, the proposed PLL model can generate a clean high frequency output a hundred times larger than the distortion input frequency of 50 MHz. The presence of jitter at VCO, ($\Phi_{VCO}$) was also implemented as shown in Fig. 6. The VCO is able to tolerate distortions generated from the jitter as illustrated in Fig. 10. This means that the proposed PLL circuit is capable of handling GHz data and removing jitter at input PD and VCO as illustrated in Fig. 5 and Fig. 10.

## 5. CONCLUSION

A simple model of PLL architecture has been demonstrated and transfer function of several important parts is described. The proposed design composed of PD, RLC low–pass filter, VCO, feedback divider and a resistor was used to realize a stable system. In this PLL, a high frequency sinusoidal signal of 5 GHz is generated with 50 MHz input signal. A simple PLL system tolerant to jitter was successfully modelled by running several configurations with different jitter inputs. It was shown through simulations using Verilog AMS that the PLL model is robust and suitable for CDR applications in the GHz frequency range.

## ACKNOWLEDGEMENT

This work is funded by the Ministry of Science and Technology's (MOSTI) Techno Fund Grant TF0409D100.

# REFERENCES

[1] B. Casper and F. O'Mahony, "Clocking analysis, implementation and measurement techniques for high-speed data links - a tutorial," *IEEE Transactions on Circuits*, vol. 56, pp. 17-39, 2009.

[2] J. Moreira and H. Wekmann, *An engineer's guide to automated testing of high-speed interfaces*. Norwood, MA: Artech House, 2010.

[3] J. Encinas, *Phase locked loops*: Chapman & Hall, 1993.

[4] M. Saber, Y. Jitsumatsu, and M. T. A. Khan, "Design and implementation of low power digital phase-locked loop," presented at Information Theory and its Applications (ISITA), 2010 International Symposium on.

[5] S. Kim, D. Lee, Y.-S. Park, Y. Moon, and D. Shim, "A dual PFD phase rotating multi-phase PLL for 5 Gbps PCI Express Gen2 Multi-Lane Serial Link Receiver in 0.13um CMOS," presented at VLSI Circuits, 2007 IEEE Smposium on, 2007.

[6] T. K. -Siang, M. S. Sulaiman, M. Reaz, C. Hean-Teik, and M. Sachdev, "A fully-integrated 5 Gbit/s CMOS clock and data recovery circuit," *Analog Integrated Circuits Signals Process (2007)*, pp. 101-109, 2007.

[7] B. Razavi, "A 10-Gb/s CMOS clock and data recovery circuit with a half-rate linear phase detector," *IEEE Journal of Solid-State Circuits*, vol. 36, pp. 761-768, 2001.

[8] T.K.Siang, M.S.Sulaiman, C.H.Teik, and M.Sachdev, "Design of high-speed clock and data recovery circuits," pp. 15-23, 2007.

[9] T.Oura, Y.Hiraku, T. Suzuki, and H.Asai, "Modeling and simulation of phase-locked loop with Verilog-A description for top-down design," presented at Circuits and Systems, 2004. Proceedings. The 2004 IEEE Asia-Pacific Conference on 2004.

[10] J.Meyer, "Modeling phase-locked loops using verilog," presented at 39th Annual precise Time and Time Interval (PTTI) Meeting, 2007.

[11] B. Razavi, *Design of Analog CMOS Integratedircuits*. Los Angeles: McGraw-Hill 2001.